\documentclass[universe,article,accept,pdftex,moreauthors]{Definitions/mdpi} 
\usepackage{psfrag}
\usepackage{graphicx} 
\usepackage{dcolumn} 
\usepackage{latexsym,amsfonts} 
\usepackage{bm} 
\usepackage{amssymb}
\usepackage{rotating}
\def\beq{\begin{equation}}
\def\eeq{\end{equation}}

\firstpage{1} 
\pubvolume{10}
\issuenum{12}
\articlenumber{460}
\pubyear{2024}
\copyrightyear{2024}
\externaleditor{Academic Editor: Andreas Fring}
\datereceived{17 September 2024 } 
\daterevised{6 December 2024 } % Comment out if no revised date
\dateaccepted{17 December 2024 } 
\datepublished{19 December 2024} 
\hreflink{https://doi.org/10.3390/\linebreak universe10120460} % If needed use \linebreak
\Title{{Spatial} Distribution of Ultracold Neutron Probability Density in the Gravitational Field of the Earth Above a Mirror}

\TitleCitation{Spatial Distribution of Ultracold Neutron Probability Density in the Gravitational Field of the Earth Above a Mirror}

\Author{{Derar} %MDPI: Please carefully check the accuracy of names and affiliations.  
 Altarawneh $^{1,}$*, {Roman}~H\"ollwieser $^{2}$\orcidA{} and {Markus} Wellenzohn $^{3,4}$}

\AuthorNames{D.~Altarawneh, R.~H\"ollwieser and M. Wellenzohn}

\AuthorCitation{{Altarawneh,} %MDPI: Please carefully check the accuracy of names
 D.; H\"ollwieser, R.; Wellenzohn, M.}
\address{%
$^{1}$ \quad Department of Applied Physics, Tafila Technical University, Tafila 66110, Jordan\\
$^{2}$ \quad Department of Theoretical Physics, Faculty of Mathematics and Natural Sciences, Wuppertal University, Gau{\ss}stra{\ss}e 20, 42119 Wuppertal, Germany %MDPI: This email address is different from the one submitted online at susy.mdpi.com(hoellwieser@uni-wuppertal.de). Please confirm which one is correct.
\\
$^{3}$ \quad Department of Engineering, Applied Electronics and Technical Informatics, University of Applied Sciences Vienna (FH Campus Wien), 1100 Vienna, Austria\\
$^{4}$ \quad Research Center IT-Security, Department of Engineering, University of Applied Sciences Vienna (FH Campus Wien), 1100 Vienna, Austria}

\corres{Correspondence: derar@ttu.edu.jo}

\abstract{We present a theoretical analysis of the experimental data reported by {Ichikawa} %MDPI: Reference citation is not allowed in Abstract. The following highlights are the same. Please revise.
 {et al.}  %(Phys. Rev. Lett. {\bf 112}, 071101 (2014)) 
on the spatial distribution of ultracold neutrons in the Earth's gravitational field above a mirror. The data involve a projection onto a pixelated detector via scattering by a cylindrical mirror.
Our study includes a calculation of the theoretical spatial distribution of the probability density associated with the quantum gravitational states of ultracold neutrons. Furthermore, we analyze this spatial distribution using the Wigner function framework.
Based on our analysis, we cannot confirm that the experimental data reported by {Ichikawa} {et al.} %(Phys. Rev. Lett. {\bf 112}, 071101 (2014)) 
correspond to the spatial distribution of quantum gravitational states of ultracold neutrons.}



\begin{document}

\section{Introduction}
\label{sec:introduction}

Quantum gravitational states of ultracold neutrons in the gravitational field of the Earth above a mirror have a long history \cite{Gibbs1975, Luschikov1978, Wallis1992, Gea1999, Abele2008} and have been experimentally observed in several studies \cite{Nesvizhevsky2002, Nesvizhevsky2003, Nesvizhevsky:2003at, Abele2003, Nesvizhevsky2005, Voronin:2005xg, Westphal:2006dj, Abele:2006xd, Naganawa2018, Muto2022}. The experimental analysis of the spatial probability density distribution for these quantum gravitational states has been extensively studied. Such investigations typically involve measuring the free fall of ultracold neutrons onto a mirror, as reviewed in \cite{Nesvizhevsky2000, Nesvizhevsky20006, AbeleWF1, AbeleWF2, BAESSLER2011, AbeleWF3}.

Furthermore, transitions between quantum gravitational states of ultracold neutrons, bouncing between two mirrors in the Earth's gravitational field, have been experimentally examined in \cite{Jenke2011, Jenke2012}. Theoretical calculations of wave functions and binding energies for quantum gravitational states of ultracold neutrons confined between two mirrors are reported in \cite{Ivanov2013} (see also \cite{Mota:2006ed, Brax:2011hb}).

Recently, experimental data on the spatial distribution of ultracold neutrons in the Earth's gravitational field above a mirror, obtained by Ichikawa {et al.} \cite{Ichikawa2014}, were interpreted as representing the spatial distribution of quantum gravitational states. In their study, the authors analyzed the vertical distribution of ultracold neutrons above a horizontal mirror, as projected onto a pixelated horizontal detector. This was achieved by scattering neutrons off a cylindrical mirror. The observed spatial modulation was interpreted as the spatial distribution of quantum gravitational states. To support this interpretation, the authors performed a theoretical analysis of the experimental data using the Wigner function formalism \cite{Wigner1932a, Wigner1932b}.

In this paper, we analyze the spatial distribution of quantum gravitational states of ultracold neutrons for the experimental setup employed by Ichikawa {et al.} \cite{Ichikawa2014}. Specifically, we examine the distribution in terms of: (i) the spatial probability density \cite{Davydov1965} and (ii) the spatial distribution of the Wigner function \cite{Wigner1932a,Wigner1932b} of these states. However, our theoretical analysis indicates that the experimental data reported by Ichikawa {et al.} \cite{Ichikawa2014} do not conclusively correspond to the spatial distribution of quantum gravitational states of ultracold neutrons.

\section{Experimental Setup of Ichikawa's Experiment on Spatial 
Distribution of Quantum Gravitational States of Ultracold Neutrons}
\label{sec:expsetup}

\begin{figure}[H]
%\centering
\includegraphics[width=0.55\linewidth]{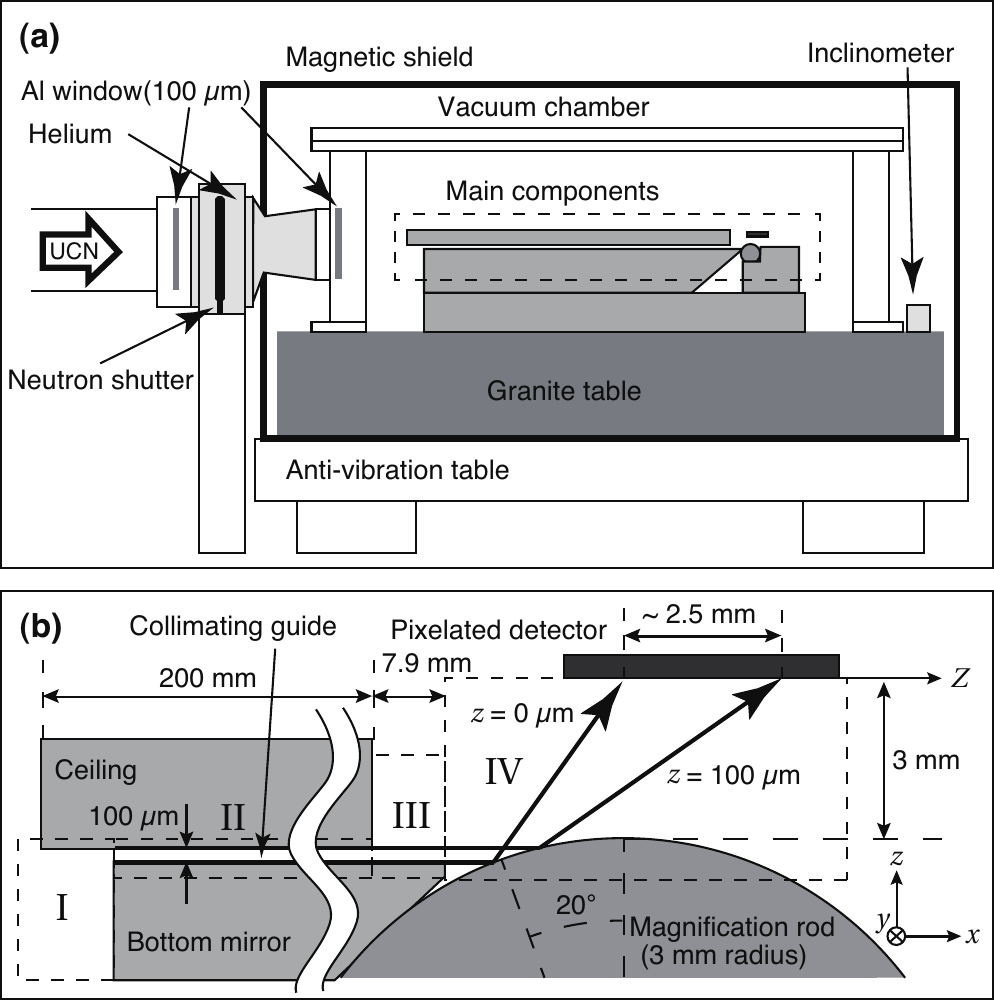}
\caption{The experimental setup for measuring the spatial distribution of ultracold neutrons in the Earth's gravitational field above a cylindrical mirror \cite{Ichikawa2014}.}
\label{fig:DN1q}
\end{figure}

The experimental setup for measuring the spatial distribution of ultracold neutrons above a cylindrical mirror, shown in Figure~\ref{fig:DN1q}, is adapted from \cite{Ichikawa2014}.  

According to \cite{Ichikawa2014}, ultracold neutrons moving in the \(x\)-direction in  spatial region II, with a length \(L_{\rm II} = 192.1\,{\rm mm}\) between two parallel plane mirrors separated by a distance \(h_{\rm II} = 0.1\,{\rm mm}\), are prepared in a quantum gravitational state \(\tilde{\psi}(z,0)\). This state is a mixed state, expressed as \(\tilde{\psi}(z,0) = \sum_{n'} \tilde{a}_{n'}(0)\,e^{i \alpha_{n'}} \tilde{\psi}_{n'}(z,0)\), where \(\tilde{\psi}_{n'}(z,0)\) are stationary pure quantum gravitational states of ultracold neutrons. Here, \(n' = 1, 2, \ldots\) represents the principal quantum number, and \(t = 0\) corresponds to the time at which ultracold neutrons are injected into spatial region II. The phases \(\alpha_{n'}\) are random.

The equal population of quantum states arises from a thermal equilibrium and the small energy differences between states at ultracold temperatures, leading to a nearly uniform distribution of neutrons among the available states. The randomness of the phases is attributed to decoherence, the preparation process, and the quantum measurement process, all of which contribute to a lack of coherence between the phases of neutrons in different quantum states.

The time evolution of ultracold neutrons in spatial region II for \(t \geq 0\) is described by the wave function \(\tilde{\psi}(z,t) = \sum_{n'} \tilde{a}_{n'}(0)\,e^{i\alpha_{n'}} \tilde{\psi}_{n'}(z,t)\). Here, \(\tilde{\psi}_{n'}(z,t) = \tilde{\psi}_{n'}(z)\,e^{-i \tilde{E}_{n'} t}\), where \(\tilde{\psi}_{n'}(z)\) are real functions, and \(\tilde{E}_{n'}\) is the binding energy of ultracold neutrons in the \(n'\)-quantum gravitational state between the two mirrors.

The horizontal motion of ultracold neutrons can, in principle, be described by a plane wave \(e^{ip_x x - i\tilde{E}_x t}\), where \(p_x = m_n v_x\) and \(\tilde{E}_x = p_x^2 / 2m_n\) represent the momentum and energy of horizontal motion, respectively. This approximation is valid, as there are no significant external forces or potential gradients affecting the neutrons horizontally, allowing them to behave as free particles in that direction. The plane wave simplification effectively represents particles with constant momentum, aligning with the conditions of horizontal motion for ultracold neutrons. This description simplifies the analysis of neutron behavior, particularly in scattering and interference experiments.

The horizontal velocity \(v_x\) of ultracold neutrons follows a nearly Gaussian distribution centered at a mean value \(v_0 = 9.4\,{\rm m/s}\) with a standard deviation \(\Delta v_x = 2.8\,{\rm m/s}\)~\cite{Ichikawa2014}. Thus, the wave function of ultracold neutrons in spatial region II is \(\tilde{\psi}_{p_x}(x,z,t) = e^{ip_x x - iE_x t} \sum_{n'}$ $\tilde{a}_{n'}(0)\,e^{i\alpha_{n'}} \tilde{\psi}_{n'}(z,t)\), with \(\sum_{n'} |\tilde{a}_{n'}(0)|^2 = 1\).

After traversing spatial region II, ultracold neutrons enter spatial region III, with a length \(L_{\rm III} = 7.9\,{\rm mm}\), bounded by a mirror below. In this region, the stationary pure quantum gravitational states are described by the wave functions \(\psi_n(z,t) = \psi_n(z)\,e^{-i E_n t}\), where \(E_n\) is the binding energy, and \(n = 1, 2, \ldots\) denotes the principal quantum number. The wave function of the mixed state \(\tilde{\psi}(z,t)\) transforms into \(\psi(z,t) = \sum_{n} a_n(t_0)\,e^{i\beta_n}\psi_n(z,t)\), where \(\beta_n\) are random phases. The coefficients \(a_n(t_0)\) are defined as:
\[
a_n(t_0) = e^{-i \beta_n} \sum_{n'} \tilde{a}_{n'}(0)\,e^{i \alpha_{n'}} \int^{h_{\rm II}}_0 \psi_n^*(z,t_0)\,\tilde{\psi}_{n'}(z,t_0)\,dz.
\]
{At} %MDPI: Please confirm if indentation should be added. The following highlights are the same.
 \(t = t_0 = L_{\rm II} / v_0\), and the transition occurs between regions II and III. Integration is performed over \(0 \leq z \leq h_{\rm II}\), as \(\tilde{\psi}_{n'}(z,t)\) vanishes outside this range. The differences in binding energies \(\Delta E_{nn'} = E_n - \tilde{E}_{n'}\) do not vanish because \(E_n \neq \tilde{E}_{n'}\) for any \(n\) and \(n'\).

Finally, ultracold neutrons arrive at spatial region IV, where they move above a cylindrical glass rod acting as a mirror \cite{Ichikawa2014}. The mechanism explaining the experimental results from Ichikawa {et al.} \cite{Ichikawa2014} is as follows: neutrons with a momentum \mbox{\(p_x = m_n v_0 = 29.5(8.8)\,{\rm meV}\)} and a de Broglie wavelength \(\lambda_n = 2\pi / p_x = 42(13)\,{\rm nm}\) scatter off the cylindrical mirror as classical particles. This scattering projects impact parameters \(b\) onto the \(Z\)-axis of the horizontal detector surface parallel to the \(x\)-axis. Since the impact parameter \(b\) is weighted by \(|\psi(b)|^2\), the specific spatial modulation observed along the \(Z\)-axis matches the results reported by Ichikawa {et al.} \cite{Ichikawa2014}.

\section{Impact Parameter of Scattering of Ultracold Neutrons as Classical
 Particles by Cylindrical Mirror}
\label{sec:impact}

Ultracold neutrons scatter off the cylindrical mirror with a momentum \( p_x = m v_0 = 29.5(8.8)\,{\rm meV} \), corresponding to a wavelength \( \lambda = 0.42(13)\,{\rm nm} \). This wavelength is significantly smaller than the characteristic scale of gravitational quantum states. Therefore, ultracold neutrons interact with the cylindrical mirror as classical particles. The spatial distribution of ultracold neutrons along the \( Z \)-axis of the pixelated detector is determined by the relationship between the impact parameter \( b \), the scattering angle \( \chi \), and the geometric parameters of the cylindrical mirror (\( R \), \( \alpha \)).

\begin{figure}[H]
%\centering
\includegraphics[height=0.33\textheight]{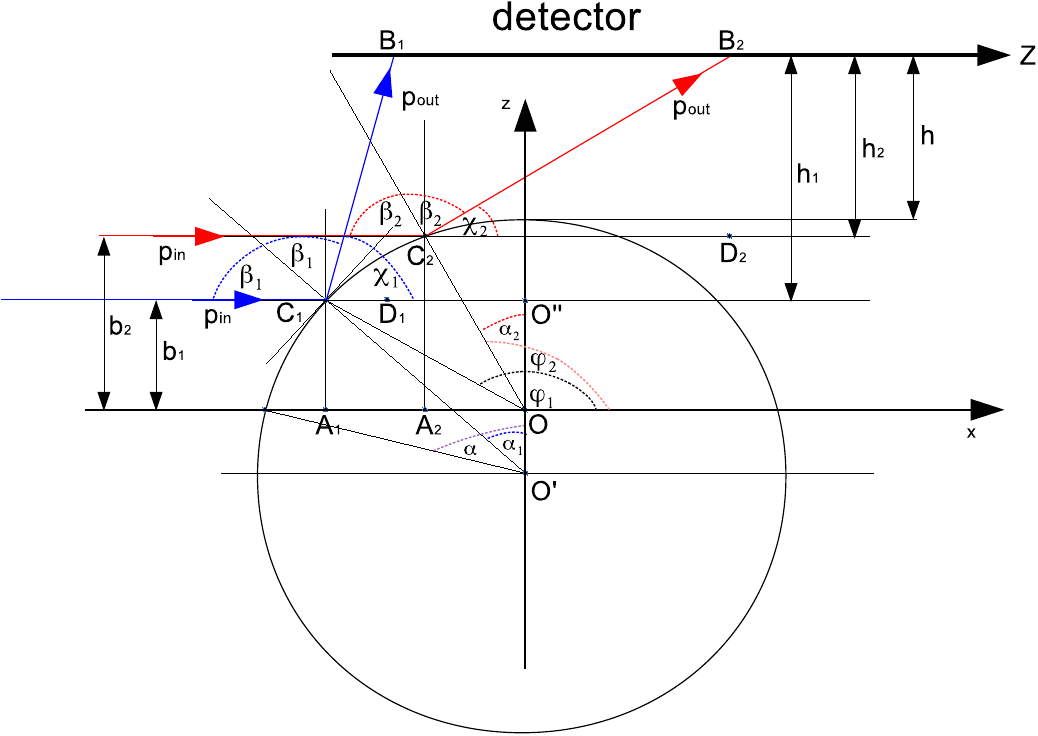}
\caption{Scattering of ultracold neutrons by a perfectly rigid cylindrical mirror as classical particles in the experiment by Ichikawa {et al.}~\cite{Ichikawa2014}.}
\label{fig:DN2qa}
\end{figure}

Figure~\ref{fig:DN2qa} illustrates the scattering geometry for ultracold neutrons interacting with a rigid cylindrical mirror, as described in the experiment by Ichikawa {et al.}~\cite{Ichikawa2014}. The radius of the mirror is \( R = 3\,{\rm mm} \), and the angle \( \alpha = \pi/9 \).

Based on the geometry shown in Figure~\ref{fig:DN2qa}, the impact parameter \( b \) can be expressed as a function of the angle \( \alpha \) and the azimuthal angle \( \varphi \) as follows:
\[
b(\alpha, \varphi) = R \sin \varphi \sqrt{1 - 2 \cos\alpha\,\cos\beta + \cos^2\alpha},
\]
where \( \beta \) is related to the scattering angle \( \chi \) by \( \beta = \chi/2 \).

The azimuthal angle \( \varphi \) and \( \beta \) are further related by:
\[
\cos \beta = \cos\alpha\,\cos^2 \varphi + \sin\varphi \sqrt{1 - \cos^2\alpha\,\cos^2 \varphi},
\]
\[
\sin \beta = \cos\varphi \left( \sqrt{1 - \cos^2\alpha\,\cos^2 \varphi} - \cos\alpha\,\sin\varphi \right).
\]
{Inverting} 
 these relationships, \( \varphi \) can be expressed as follows:
\[
\cos \varphi = -\frac{\sin \beta}{\sqrt{1 - 2 \cos\alpha\,\cos\beta + \cos^2\alpha}},
\]
\[
\sin \varphi = +\frac{\cos \beta - \cos\alpha}{\sqrt{1 - 2 \cos\alpha\,\cos\beta + \cos^2\alpha}}.
\]

Using these expressions, the impact parameter \( b \) in terms of the scattering angle \( \chi \)~becomes:
\[
b(\alpha, \chi) = R \left( \cos \frac{\chi}{2} - \cos\alpha \right).
\]

\textls[-25]{The maximal impact parameter \( b_{\text{max}} \) corresponds to the height of region II, \mbox{\( h_{\rm II} = 0.1\,{\rm mm} \),}} which defines the minimal scattering angle:
\[
\chi_{\text{min}} = 2\arccos\left(\cos\alpha + \frac{h_{\rm II}}{R}\right) = 0.466 \, \text{rad} \, \, (\chi_{\text{min}} = 26.676^\circ).
\]
{The} 
 maximal scattering angle \( \chi_{\text{max}} \) is determined by the condition \( b(\alpha, \chi_{\text{max}}) = 0 \), giving the following:
\[
\chi_{\text{max}} = 2\alpha = \frac{2\pi}{9} \, \, (\chi_{\text{max}} = 40^\circ).
\]

Thus, the endpoints of the interval \( B_1B_2 \) on the \( Z \)-axis are shifted with respect to  points \( C_1 \) and \( C_2 \), which correspond to the maximal and minimal scattering angles, respectively. These shifts are calculated as follows:
\[
C_1D_1 = \left(h + R(1 - \cos\alpha)\right) \cot\chi_{\text{max}} = 3.79\,{\rm mm},
\]
\[
C_2D_2 = \left(h + R(1 - \cos\alpha) - h_{\rm II}\right) \cot\chi_{\text{min}} = 6.13\,{\rm mm}.
\]

Points \( D_1 \) and \( D_2 \) are the projections of \( B_1 \) and \( B_2 \), respectively. As a result, the spatial distribution of ultracold neutrons can be observed along the \( Z \)-axis in the following interval:
\[
B_1B_2 = 2.34\,{\rm mm}.
\]

\vspace{-6pt}

\section{Projection of Ultracold Neutrons with Impact Parameter \boldmath$b$
 onto \boldmath$Z$-Axis of \mbox{Pixelated Detector}}
\label{sec:projection}

As we have shown in Section~\ref{sec:impact}, ultracold neutrons,
moving with an impact parameter $b(\alpha, \chi)$ with a velocity
$v_0$, are projected by a cylindrical mirror onto the $Z$-axis of the
pixelated detector. We may define such a projection as follows:
\begin{eqnarray}\label{eq:6}
\hspace{-0.3in}Z = \frac{h + R(1 - \cos\alpha) - b(\alpha,
  \chi)}{\tan\chi} - \frac{h + R(1 - \cos\alpha)}{\tan\chi_{\rm max}},
\end{eqnarray}
where $Z = 0$ for $b = 0$. Using the relations
\begin{eqnarray}\label{eq:7}
\hspace{-0.3in}\tan\chi = \frac{\displaystyle \Big(\cos\alpha +
  \frac{b}{R}\Big)\sqrt{\sin^2\alpha - 2\cos \alpha\,\frac{b}{R} -
    \frac{b^2}{R^2}}}{\displaystyle \frac{b^2}{R^2} + 2\cos
  \alpha\,\frac{b}{R} + \frac{1}{2}\,\cos 2\alpha}\quad,\quad
\tan\chi_{\rm max} = \tan 2\alpha,
\end{eqnarray}
{we} 
 obtain $Z$ as a function of an impact parameter $b$
 \vspace{-12pt}
\begin{adjustwidth}{-\extralength}{0cm}
%\centering %% If there is a figure in wide page, please release command \centering
\begin{eqnarray}\label{eq:8}
\hspace{-0.3in}Z = \frac{h + R(1 - \cos\alpha) - b}{\displaystyle
  \Big(\cos\alpha + \frac{b}{R}\Big)\sqrt{\sin^2\alpha - 2\cos
    \alpha\,\frac{b}{R} - \frac{b^2}{R^2}}}\Bigg(\frac{b^2}{R^2} +
2\cos \alpha\,\frac{b}{R} + \frac{1}{2}\,\cos 2\alpha\Bigg) - \Big(h +
R(1 - \cos\alpha)\Big)\cot 2\alpha.
\end{eqnarray}
\end{adjustwidth}
{At} 
 $ b = 0$ we obtain $Z = 0$. Then, taking into account the values of
the parameters of the experimental setup of Ichikawa's experiment
\cite{Ichikawa2014}, we may approximate the impact parameter $b$ by the
expression
\begin{eqnarray}\label{eq:9}
b(Z) = R\,\frac{\sin^2\alpha}{2 \cos\alpha}\,\frac{Z(Z + 2
  Z_0)}{Z^2_0}\,\frac{1}{\displaystyle \frac{\sin^2\alpha(2 +
    4\cos^2\alpha - 4\cos^3\alpha)}{\cos^2\alpha (2 - \cos\alpha) \cos
    2\alpha} + \frac{(Z + Z_0)^2}{Z^2_0}}
\end{eqnarray}
where $Z_0 = 3.79\,{\rm mm}$. At $Z = 2.34\,{\rm mm}$, we obtain $b =
0.101\,{\rm mm}$. Thus, Equation~(\ref{eq:9}) allows to fit the maximal
value of the impact parameter with an accuracy of about $1\%$. A
derivative $db(Z)/dZ$ is
\begin{eqnarray}\label{eq:10}
\frac{db(Z)}{d Z} = \frac{\sin^2\alpha}{\cos\alpha}\,\frac{R(Z +
  Z_0)}{Z^2_0}\frac{\displaystyle 1 + \frac{\sin^2\alpha(2 +
    4\cos^2\alpha - 4\cos^3\alpha)}{\cos^2\alpha (2 - \cos\alpha) \cos
    2\alpha}}{\displaystyle \Bigg(\frac{\sin^2\alpha(2 + 4\cos^2\alpha -
    4\cos^3\alpha)}{\cos^2\alpha (2 - \cos\alpha) \cos 2\alpha} +
  \frac{(Z + Z_0)^2}{Z^2_0}\Bigg)^2}.
\end{eqnarray}
{Now,} 
 we are able to use the impact parameter $b(Z)$ for the analysis of
a spatial distribution of ultracold neutrons by a pixelated detector.

\section{Spatial Distribution of Energy Levels of Quantum Gravitational
 States of Ultracold Neutrons Between Two Mirrors}
\label{sec:quasiclassical}

For the subsequent analysis of the experimental data by Ichikawa {et al.} \cite{Ichikawa2014}, it is necessary to determine the distribution of quantum gravitational states of ultracold neutrons within spatial region II. The energy levels of these states are defined by the roots of the following equation \cite{Ivanov2013}:

\begin{eqnarray}\label{eq:11}
\hspace{-0.3in}{\rm Ai}(\tilde{\xi}_{n'}){\rm Bi}\Big(\tilde{\xi}_{n'}
+ \frac{h_{\rm II}}{\ell_0}\Big) - {\rm Ai}\Big(\tilde{\xi}_{n'} +
\frac{h_{\rm II}}{\ell_0}\Big){\rm Bi}(\tilde{\xi}_{n'}) = 0,
\end{eqnarray}
where \( h_{\rm II} = 0.1\,{\rm mm} \) is the height of spatial region II, and \( \ell_0 = (2m^2g)^{-1/3} = 5.87\,{\rm \upmu m} \) {is} %MDPI: We revised unit. Please confirm.
 a quantum scale associated with the quantum gravitational states of ultracold neutrons. Here, \( g \) represents the gravitational acceleration. The roots \( \tilde{\xi}_{n'} \) of Equation~(\ref{eq:11}) define the energy levels \( \tilde{E}_{n'} = - m g \ell_0 \tilde{\xi}_{n'} \), where \( n' = 1, 2, \ldots \) is the principal quantum number, and \( m g\ell_0 = 0.602\,{\rm peV} \).

The maximal number of quantum gravitational states \( n'_{\rm max} \) is determined by the condition \( -\tilde{\xi}_{n'_{\rm max}} \le h_{\rm II}/\ell_0 \). Setting \( -\tilde{\xi}_{n'_{\rm max}} = h_{\rm II}/\ell_0 \) and recognizing that \( {\rm Ai}(0) = {\rm Bi}(0) \), we arrive at the equation. For \( n'_{\rm max} \gg 1 \) and using the asymptotic behavior of the Airy functions for \( -\tilde{\xi}_{n'_{\rm max}} \gg 1 \), Equation~(\ref{eq:11}) can be rewritten as follows:

\begin{eqnarray}\label{eq:12}
\hspace{-0.3in}\sin\Big(\frac{2}{3}\,(- \tilde{\xi}_{n'_{\rm
    max}})^{3/2}\Big) = 0,
\end{eqnarray}
{the} 
 root of which is given by:

\begin{eqnarray}\label{eq:13}
\hspace{-0.3in} \pi n'_{\rm max} = \frac{2}{3}\,(-
\tilde{\xi}_{n'_{\rm max}})^{3/2} \le \frac{2}{3}\,\Big(\frac{h_{\rm
    II}}{\ell_0}\Big)^{3/2}.
\end{eqnarray}

This defines the maximal number \( n'_{\rm max} = (2/3\pi)(h_{\rm
  II}/\ell_0)^{3/2} \) of the quantum gravitational states of ultracold neutrons within spatial region II.

The same result, along with the spatial distribution of quantum gravitational levels in region II, can be obtained using the quasi-classical approximation of quantum mechanics~\cite{Davydov1965}. In this approximation, the maximal number of quantum gravitational states of ultracold neutrons in the spatial region \( 0 \le z \le h_{\rm II} \) is given by:

\begin{eqnarray}\label{eq:14}
\hspace{-0.3in} n'_{\rm max} \le \int^{h_{\rm II}}_0 \int^{p(z)}_0
\frac{dp dz}{\pi} = \frac{1}{\pi}\int^{h_{\rm II}}_0 p(z)dz =
\frac{2}{3\pi}\,\Big(\frac{h_{\rm II}}{\ell_0}\Big)^{3/2},
\end{eqnarray}
where \( p(z) = \sqrt{2 m g^2 (h_{\rm II} - z)} = \sqrt{h_{\rm II} - z}/\ell^{3/2} \) is the classical momentum of ultracold neutrons. 

Using Equation~(\ref{eq:14}), we can determine the spatial distribution of quantum gravitational levels in region II. This yields:

\begin{eqnarray}\label{eq:15}
\hspace{-0.3in} \frac{dn'(z)}{dz} = \frac{1}{\pi
  \ell^{3/2}_0}\,\sqrt{h_{\rm II} - z}.
\end{eqnarray}

The probability distribution of quantum gravitational states within the spatial region \( 0 \le z \le h_{\rm II} \) is expressed as follows:

\begin{eqnarray}\label{eq:16}
\hspace{-0.3in} \frac{dP(z)}{dz} = \frac{3}{2}\,\frac{1}{h^{3/2}_{\rm
    II}}\,\sqrt{h_{\rm II} - z},
\end{eqnarray}
where \( P(z) = n'(z)/n'_{\rm max} \). The probability \( P(z \le h) \) of finding quantum gravitational states within the spatial region \( 0 \le z \le h \) is given by:

\begin{eqnarray}\label{eq:17}
\hspace{-0.3in} P(z \le h) = \int^h_0 \frac{dP(z)}{dz}\,dz = 1 - \Big(1 - \frac{h}{~h_{\rm II}}\Big)^{3/2}.
\end{eqnarray}

These results allow us to calculate the expansion coefficients of the wave function for the mixed state of ultracold neutrons in spatial region II.

In practice, these findings indicate that the spatial distribution of ultracold neutrons between two mirrors is entirely determined by their phase volume. This is further supported by treating ultracold neutrons as an ideal non-relativistic classical gas, confined between two mirrors in the spatial region \( 0 \le z \le h_{\rm II} \), with a Maxwell--Boltzmann distribution function \( f(p_z, z) \) in the phase volume at temperature \( T \) \cite{Landau1979}. The Maxwell--Boltzmann distribution function \( f(p_z, z) \) is normalized to the total number of ultracold neutrons \( N \):

\begin{eqnarray}\label{eq:19}
N = \int^{h_{\rm II}}_0\int_{E\ge \frac{p^2_z}{2m} +
  mgz}f(p_z,z)\,\frac{dp_z dz}{2\pi}.
\end{eqnarray}

For ultracold neutrons where \( T \gg E \), \( \exp(-E/T) \) can be approximated by unity, simplifying the calculation yields:

\begin{eqnarray}\label{eq:19a}
N \propto 2\int^{h_{\rm II}}_0\int^{p(z)}_0\,\frac{dp_z dz}{2\pi} =
\frac{1}{\pi} \int^{h_{\rm II}}_0 p(z)\,dz,
\end{eqnarray}
where \( p(z) = \sqrt{2m(E - mgz)} \). With energy conservation and \( p(h_{\rm II}) = 0 \) at \( z = h_{\rm II} \), the result is:

\begin{eqnarray}\label{eq:19b}
N \propto \frac{1}{\pi \ell^{3/2}_0}\int^{h_{\rm II}}_0\sqrt{h_{\rm
    II} - z}\,dz = \frac{2}{3\pi}\,\Big(\frac{h_{\rm
    II}}{\ell_0}\Big)^{3/2}.
\end{eqnarray}

This confirms the well-known result \( N \propto h^{3/2}_{\rm II} \) \cite{Ruess2000}. The spatial distribution of ultracold neutrons between two mirrors, normalized to the total number \( N \), is given by:

\begin{eqnarray}\label{eq:19c}
\frac{d N(z)}{d z} = N\,\frac{3}{2}\,\frac{1}{h^{3/2}_{\rm
    II}}\,\sqrt{h_{\rm II} - z}.
\end{eqnarray}

The total number \( N(z \le h) \) in the spatial region \( 0 \le z \le h \) is expressed as follows:

\begin{eqnarray}\label{eq:19d}
N(z \le h) = \int^h_0 \frac{d N(z)}{d z}\,dz = N\Big(1 - \Big(1 -
\frac{h}{~h_{\rm II}}\Big)^{3/2}\Big).
\end{eqnarray}

The spatial distribution of ultracold neutrons, as described in Equation~(\ref{eq:19c}), matches the \( z \)-dependence of the spatial distribution of quantum gravitational energy levels.

\section{Spatial Distribution of Probability Density of Quantum 
Gravitational States of Ultracold Neutrons}
\label{sec:probability}

According to \cite{Gibbs1975}, ultracold neutrons moving in the gravitational field of the Earth above a mirror can exist in quantum gravitational states described by the wave functions \(\psi_{n}(z)\), which are expressed as follows \cite{Gibbs1975}:

\[
\psi_{n}(z) = \frac{\displaystyle \text{Ai}\left(\xi_{n} + \frac{z}{\ell_0}\right)}{\displaystyle \sqrt{\int_{0}^\infty dz\,\Big|\text{Ai}\Big(\xi_{n} + \frac{z}{\ell_0}\Big)\Big|^2}} = \frac{1}{\sqrt{\ell_0}}\,\frac{\displaystyle \text{Ai}\Big(\xi_{n} + \frac{z}{\ell_0}\Big)}{\displaystyle \sqrt{\int_{0}^\infty d\xi\,|\text{Ai}(\xi_{n} + \xi)|^2}},
\]
where \(\xi = z / \ell_0\), \(\ell_0 = (2 m^2 g)^{-1/3} \approx 5.87 \times 10^{-3}\,\text{mm}\), and \(g\) is the gravitational acceleration. The roots of the Airy function, defined as \(\text{Ai}(\xi_{n}) = 0\), determine the energy spectrum of the quantum gravitational states as \(E_{n} = -m g \ell_0 \xi_{n}\) in region III for \(n = 1, 2, \ldots\), where \(m g \ell_0 = 0.602\,\text{peV}\). The spatial probability density distribution of ultracold neutrons in a quantum gravitational state with principal quantum number \(n\) is given by \(|\psi_{n}(z)|^2\). Alternatively, expressed in terms of the impact parameter \(b\), the distribution is \(|\psi_{n}(b)|^2\), where \(b\) depends on \(Z\), i.e., \(b(Z)\) (see Equation~(\ref{eq:9})).

Consequently, the spatial distribution of the probability density of ultracold neutrons along the \(Z\)-axis in the \(n'\)-th quantum gravitational state is given by:

\[
\frac{dW_{n}(Z)}{dZ} = |\psi_{n}(b(Z))|^2\,\frac{db(Z)}{dZ}.
\]

This spatial distribution is defined within the interval \(0 \leq Z \leq 2.34\,\text{mm}\). It can be shown that the integrated probability density over this range equals unity for the first 12 quantum gravitational states. For higher excited states, the probability decreases as \(n'\)~increases.

The spatial distribution of quantum gravitational states of ultracold neutrons, as measured by Ichikawa {et al.} \cite{Ichikawa2014}, depends significantly on the wave functions of quantum gravitational states in region II. Two possible constructions of these wave functions are explored below.

As shown in \cite{Ivanov2013}, the wave function of the \(n'\)-th quantum gravitational state of ultracold neutrons confined within \(0 \leq z \leq h_{\text{II}}\) (between two mirrors) is:

\[
\tilde{\psi}_{n'}(z) = \frac{1}{\sqrt{\ell_0}}\frac{\displaystyle \text{Ai}\Big(\tilde{\xi}_{n'} + \frac{z}{\ell_0}\Big)\text{Bi}(\tilde{\xi}_{n'}) - \text{Ai}(\tilde{\xi}_{n'})\text{Bi}\Big(\tilde{\xi}_{n'} + \frac{z}{\ell_0}\Big)}{\displaystyle \sqrt{\int_{0}^{h_{\text{II}}/\ell_0} \Big|\text{Ai}(\tilde{\xi}_{n'} + \xi)\text{Bi}(\tilde{\xi}_{n'}) - \text{Ai}(\tilde{\xi}_{n'})\text{Bi}(\tilde{\xi}_{n'} + \xi)\Big|^2 d\xi}}.
\]

This wave function satisfies the boundary conditions \(\tilde{\psi}_{n'}(0) = \tilde{\psi}_{n'}(h_{\text{II}}) = 0\). The roots \(\tilde{\xi}_{n'}\) of Equation~(\ref{eq:11}) define the energy spectrum \(\tilde{E}_{n'} = -m g \ell_0 \tilde{\xi}_{n'}\) for \(n' = 1, 2, \ldots\) in region II. For \(h_{\text{II}} = 0.10\,\text{mm}\), only the first 15 quantum gravitational states are allowed. Ultracold neutrons exist in a mixed state, which is a superposition of these 15 states. 

\begin{figure}[H]
%\centering 
\includegraphics[width=.52\textwidth]{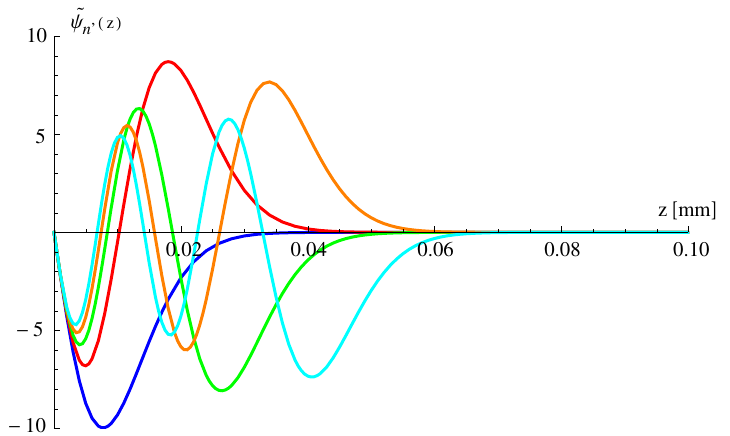}
  \includegraphics[width=.474\textwidth]{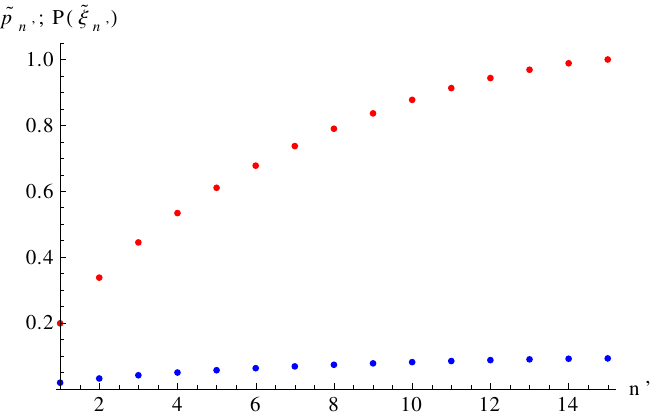}
 \caption{{(\textbf{left})} %MDPI: Please confirm if the explanation of the colors needs to be added in the figure caption.
 Wave functions of first five quantum gravitational states of
   ultracold neutrons in the spatial region $0 \le z \le h_{\rm II}$.
(\textbf{right}) The probabilities $\tilde{p}_{n'} =
|\tilde{a}_{n'}(0)|^2$ (blue dots) and
   $P(\tilde{\xi}_{n'})$ (red dots) to find the $n'$--quantum
   gravitational state of ultracold neutrons in  spatial region II.}
   \label{fig:DN5qa}
\end{figure}

For instance, Figure~\ref{fig:DN5qa} illustrates the wave functions of the first five quantum gravitational states in region II. These wave functions are highly sensitive to the precision of the roots of Equation~(\ref{eq:11}), and their computation requires precise numerical values of these~roots.

The coefficients of the wave function expansion in region III are determined by:

\[
a_{n}(t_0) = e^{-i \beta_{n}} \sum_{n'=1}^{15} \tilde{a}_{n'}(0)\,e^{i\alpha_{n'}}\,e^{i(E_{n} - \tilde{E}_{n'})t_0}\,\langle\psi_{n}|\tilde{\psi}_{n'}\rangle,
\]
where \(t_0 = L_{\text{II}} / v_0\) is the time at which neutrons transition between regions II and III, and \(\langle\psi_{n}|\tilde{\psi}_{n'}\rangle = \int_{0}^{h_{\text{II}}} dz\,\psi_{n}(z) \tilde{\psi}_{n'}(z)\).

Upon averaging over random phases, the spatial probability density distribution in region III becomes:

\[
\frac{dW(Z)}{dZ} = \sum_{n=1}^\infty p_{n}\,|\psi_{n}(b(Z))|^2\,\frac{db(Z)}{dZ},
\]
where \(p_{n}\) is the probability of finding the system in the \(n\)-th state, defined as follows:

\[
p_{n} = \sum_{n'=1}^{15} |\tilde{a}_{n'}(0)|^2 |\langle\psi_{n}|\tilde{\psi}_{n'}\rangle|^2.
\]

\begin{figure}[H]
\centering \includegraphics[width=0.484\linewidth]{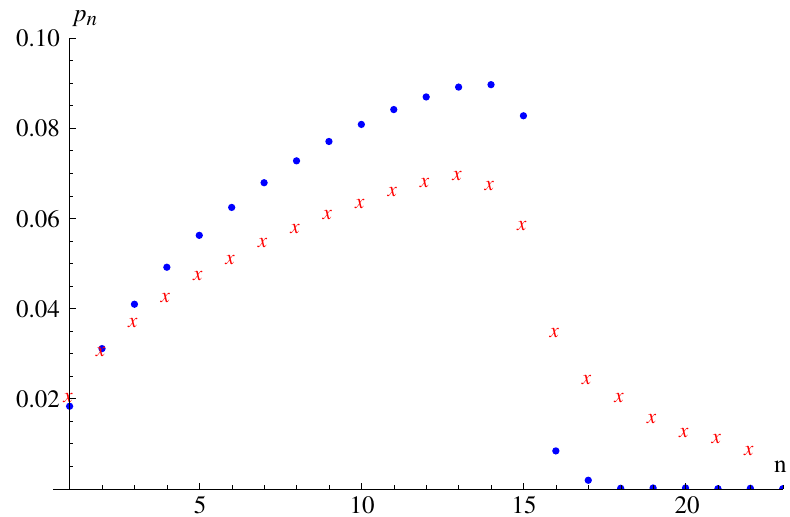}
\includegraphics[width=0.51\linewidth]{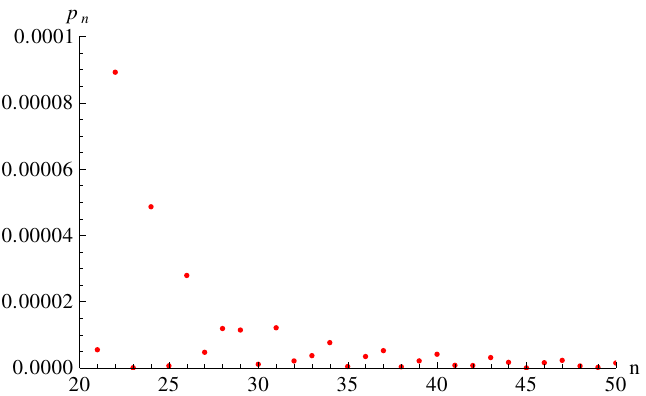}
\caption{The probabilities $p_n$ (blue dots) to find a $n$--quantum
  gravitational state in the mixed state with $n = 1,2,\ldots, 100$
  (\textbf{left}), compared to the probabilities from the paper by Ichikawa
  {et al.}  \cite{Ichikawa2014} (red crosses). The probabilities
  $p_n$ (\textbf{right}) show random oscillations for $n > 20$, which were not
  obtained by Ichikawa {et al.}  \cite{Ichikawa2014}.}
\label{fig:DN7qa}
\end{figure}

A comparison between the theoretical probabilities \(p_{n}\) and those obtained experimentally by Ichikawa {et al.} reveals significant discrepancies beyond the first two states (see {Figure} %MDPI: We revised citation. Please confirm.
 \ref{fig:DN7qa}).
The experimental data from Ichikawa {et al.} \cite{Ichikawa2014} show discrepancies with the spatial distribution of quantum gravitational states predicted theoretically. This conclusion is supported by further analysis using the Wigner function (see Section~\ref{sec:wfunction}). 

\section{{Wigner Function for Quantum Gravitational States of 
Ultracold Neutrons}}
\label{sec:wfunction}

\begin{figure}[H]
\centering 
\includegraphics[width=0.495\linewidth]{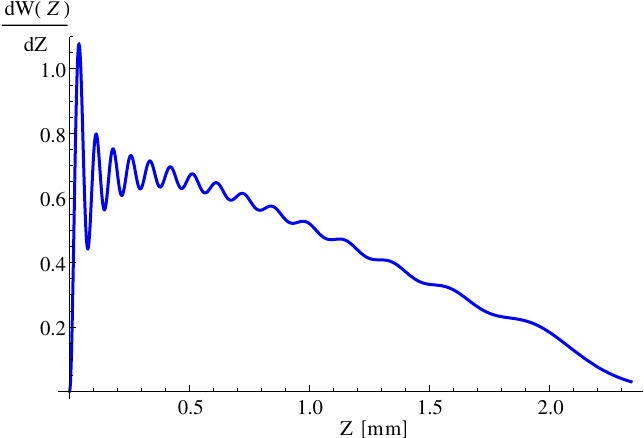}
\includegraphics[width=0.495\linewidth]{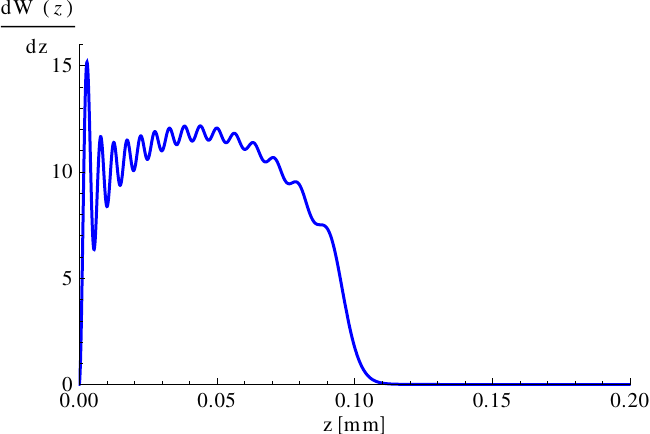}
\caption{The spatial probability density of quantum gravitational states after scattering by the cylindrical mirror and detected by the detector in the region \( 0 \le Z \le 2.34\,{\rm mm} \), demonstrating the coherence properties encoded in the Wigner function (\textbf{left}). The spatial distribution of the probability density of quantum gravitational states of ultracold neutrons (\textbf{right}) in  spatial region III before scattering by a cylindrical mirror, highlighting the gravitational quantization effects.}
\label{fig:DN8qa}
\end{figure}

In this section, we analyze the spatial distribution of quantum gravitational states of ultracold neutrons using the Wigner function \cite{Wigner1932a,Wigner1932b}, as performed by Ichikawa {et al.} \cite{Ichikawa2014}. The Wigner function for the quantum gravitational states of ultracold neutrons in region III is defined by \cite{Ichikawa2014}:

\begin{eqnarray}\label{eq:29}
  \hspace{-0.3in}W(z,p_z) =
  \int^{+\infty}_{-\infty} d\xi \, \psi^*\left(z - \frac{1}{2} \xi\right) \psi\left(z + \frac{1}{2} \xi\right) e^{-i p_z \xi},
\end{eqnarray}
where \( p_z = mv_x \sin\chi \). The Wigner function $W(z,p_z)$ offers a quasi-probabilistic representation of quantum states, describing their coherence in both position and momentum space. The dependence on $]\sin\chi(Z)$ arises from the geometrical scattering conditions and relates the vertical spatial distribution to the horizontal velocity component. By substituting \( z \to b(Z) \) (see Equation~(\ref{eq:11})) and \( \sin\chi \to \sin\chi(Z) \), where \( \sin\chi(Z) \) is given by:

\begin{eqnarray}\label{eq:30}
  \hspace{-0.3in}\sin\chi(Z) = 2\left(\cos\alpha + \frac{b(Z)}{R}\right)\sqrt{\sin^2\alpha - 2 \cos\alpha\,\frac{b(Z)}{R}},
\end{eqnarray}
{and} 
 using the expansion coefficients \( \tilde{a}_{n'}(t_0) \) and wave functions \( \tilde{\psi}_{n'}(z) \), we rewrite the Wigner function in the following form:
\vspace{-9pt}

\begin{adjustwidth}{-\extralength}{0cm}
%\centering %% If there is a figure in wide page, please release command \centering
\begin{eqnarray}\label{eq:26}
  \hspace{-0.6in}&& W(Z,p_z) = \sum^{\infty}_{n' = 1} |\tilde{a}_{n'}(t_0)|^2
  \int^{+\infty}_{-\infty} d\xi \, \tilde{\psi}_{n'}\left(b(Z) - \frac{1}{2}\xi\right) \tilde{\psi}_{n'}\left(b(Z) + \frac{1}{2}\xi\right) \cos(m v_0 \xi \sin\chi(Z)) \nonumber \\
  \hspace{-0.3in}&& + \sum_{m' > n' = 1} \text{Re} \left[ \tilde{a}^*_{m'}(t_0) \tilde{a}_{n'}(t_0) \right] \int^{+\infty}_{-\infty} d\xi
  \left[ \tilde{\psi}_{m'}\left(b(Z) - \frac{1}{2}\xi\right) \tilde{\psi}_{n'}\left(b(Z) + \frac{1}{2}\xi\right) + \tilde{\psi}_{m'}\left(b(Z) + \frac{1}{2}\xi\right) \tilde{\psi}_{n'}\left(b(Z) - \frac{1}{2}\xi\right) \right] \nonumber \\
  \hspace{-0.3in}&& \times \cos(m v_0 \xi \sin\chi(Z)) \nonumber \\
  \hspace{-0.3in}&& + \sum_{m' > n' = 1} \text{Im} \left[ \tilde{a}^*_{m'}(t_0) \tilde{a}_{n'}(t_0) \right] \int^{+\infty}_{-\infty} d\xi
  \left[ \tilde{\psi}_{m'}\left(b(Z) - \frac{1}{2}\xi\right) \tilde{\psi}_{n'}\left(b(Z) + \frac{1}{2}\xi\right) - \tilde{\psi}_{m'}\left(b(Z) + \frac{1}{2}\xi\right) \tilde{\psi}_{n'}\left(b(Z) - \frac{1}{2}\xi\right) \right] \nonumber \\
  \hspace{-0.3in}&& \times \sin(m v_0 \xi \sin\chi(Z)),
\end{eqnarray}
\end{adjustwidth}
where \( b(Z) \) is defined by Equation~(\ref{eq:11}), and the coefficients \( |\tilde{a}_{n'}(t_0)|^2 \) and \( \text{Re}\left[ \tilde{a}^*_{m'}(t_0) \tilde{a}_{n'}(t_0) \right] \) are given by {Equation} %MDPI: We revised citation. Please confirm.
 (\ref{eq:27}). The coefficients \( \text{Im}\left[ \tilde{a}^*_{m'}(t_0) \tilde{a}_{n'}(t_0) \right] \), averaged over random phases \( \alpha_n \), are as follows:

\vspace{-12pt}
\begin{adjustwidth}{-\extralength}{0cm}
%\centering %% If there is a figure in wide page, please release command \centering
\begin{eqnarray}\label{eq:27}
\text{Im}\left[ \tilde{a}^*_{m'}(t_0) \tilde{a}_{n'}(t_0) \right] =
 \sin\left( \left(\tilde{E}_{m'} - \tilde{E}_{n'}\right) \frac{L_{\rm II}}{v_0} \right) \sum^{15}_{n = 1} |a_n(0)|^2 \int^{h_{\rm II}}_0 dz' \, \tilde{\psi}_{m'}(z') \psi_n(z') \int^{h_{\rm II}}_0 dz \, \tilde{\psi}_{n'}(z) \psi_n(z),
\end{eqnarray}
\end{adjustwidth}
where \( m' > n' \), \( h_{\rm II} = 0.1\,{\rm mm} \), \( L_{\rm II} = 192.1\,{\rm mm} \), and \( v_0 = 9.4\,{\rm m/s} \). The Wigner function given by Equation~(\ref{eq:26}) is plotted in Figure~\ref{fig:DN8qa}.

After integrating the Wigner function (Equation~(\ref{eq:26})) over  velocities \( v_x \) using a Gaussian distribution, we obtain the following expression:

\vspace{-9pt}
\begin{adjustwidth}{-\extralength}{0cm}
%\centering %% If there is a figure in wide page, please release command \centering
\begin{eqnarray}\label{eq:28}
 \overline{W(z,p_z)} =
  \int^{+\infty}_{-\infty} d\xi \, \tilde{\psi}^*\left(z - \frac{1}{2} \xi\right) \tilde{\psi}\left(z + \frac{1}{2} \xi\right) \exp\left(- \frac{1}{2} \Delta v^2 m^2 \xi^2 \sin^2\chi \right) \cos\left(m v_0 \xi \sin\chi \right),
\end{eqnarray}
\end{adjustwidth}
where \( \Delta v = 2.8\,{\rm m/s} \) is the variance in the horizontal velocities of ultracold neutrons.The Gaussian approximation for the horizontal velocity distribution $v_x$ reflects the experimental observations of ultracold neutron sources, where horizontal velocities exhibit a thermal spread. The contributions of the crossing terms \( \text{Re}\left[a^*_{m'}(t_0) a_{n'}(t_0)\right] \) and \( \text{Im}\left[a^*_{m'}(t_0) a_{n'}(t_0)\right] \) are negligible due to strong oscillations and the smallness of their amplitudes.
Numerical simulations confirm that the oscillatory crossing terms  average out over the integration range due to their high-frequency behavior. This ensures their contributions to $W(Z,p_z)$ are negligible relative to the diagonal terms.

\begin{figure}[H]
 \includegraphics[width=0.49\linewidth]{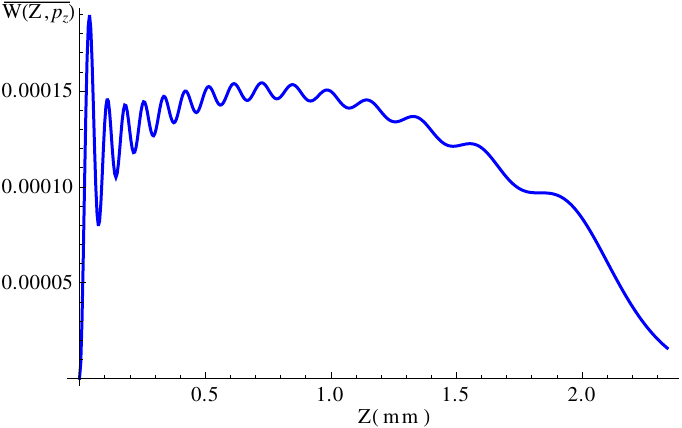}
\includegraphics[width=0.5\linewidth]{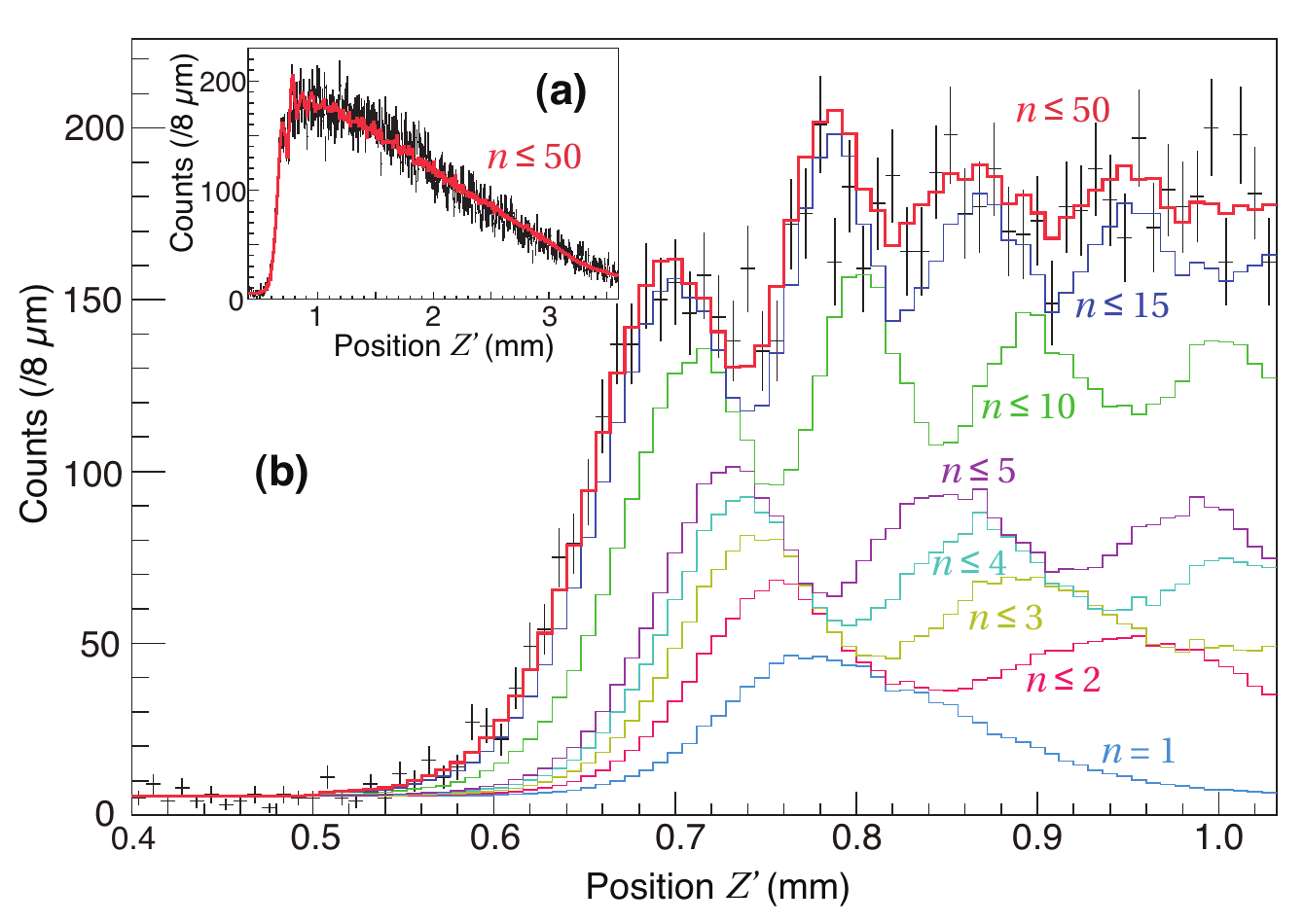}
\caption{A spatial distribution of the Wigner function $\overline{W(Z,p_z)}$
  (\textbf{left}) of quantum gravitational states of ultracold neutrons,
  scattered by a cylindrical mirror and detected by a pixelated
  detector. The distributions of ultracold neutrons (\textbf{right}) at a
  pixelated detector for the data (points and errors) and for the
  prediction with the best fit parameters (lines) for the whole region
  (\textbf{a}) and the lower $Z'$--region (\textbf{b}). $n \le 50$ eigenstates are
  considered in the calculation. The distributions with the selected
  eigenstates are also shown in (\textbf{b}) (taken from
  \cite{Ichikawa2014}). }
\label{fig:DN9q}
\end{figure}

By replacing \( z \to b(Z) \) and using Equation~(\ref{eq:11}), we obtain the spatial distribution of quantum gravitational states of ultracold neutrons in a pixelated detector in terms of the Wigner function \( \overline{W(z,p_z)} \):
\vspace{-9pt}

\begin{eqnarray}\label{eq:29}
  \hspace{-0.3in}\overline{W(Z,p_z)} &=&
  \int^{+\infty}_{-\infty}d\xi\,\sum^{\infty}_{n' =
    1}|\tilde{a}_{n'}(t_0)|^2\tilde{\psi}_{n'}\Big(b(Z) -
  \frac{1}{2}\,\xi\Big)\tilde{\psi}_{n'}\Big(b(Z) +
  \frac{1}{2}\,\xi\Big)\nonumber\\
 \hspace{-0.3in}&&\times\, \exp\Big(- \frac{1}{2}\,\Delta v^2
 m^2\xi^2\,\sin^2\chi(Z)\Big)\,\cos\Big(mv_0\xi\,\sin\chi(Z)\Big).
\end{eqnarray}
{In} 
 Figure~\ref{fig:DN9q} we plot the Wigner function
$\overline{W(Z,p_z)}$ in the interval $0 \le Z \le 2.34\,{\rm mm}$. 

\iffalse
\begin{eqnarray}\label{eq:31}
  \hspace{-0.3in}W(Z,p_z) = \sum^{\infty}_{n = 1} p_{n}
  \int^{+\infty}_{-\infty}d\xi\,\psi_{n}\Big(b(Z) -
  \frac{1}{2}\,\xi\Big) \psi_{n}\Big(b(Z) +
  \frac{1}{2}\,\xi\Big)\,\cos(m v_x \xi \sin\chi(Z)),
\end{eqnarray}
where $b(Z)$ is defined by Equation~(\ref{eq:11}) and the probabilities
$p_{n}$ are given by Equation~(\ref{eq:26})). For $v_x = v_0$ the Wigner
function Equation~(\ref{eq:31}) is plotted in Figure\,\ref{fig:DN9q}.
\fi

\section{Discussion}

The results presented in this work highlight significant discrepancies between the theoretical predictions and  experimental data on the spatial distribution of quantum gravitational states of ultracold neutrons obtained by Ichikawa {et al.} \cite{Ichikawa2014}. While the theoretical framework accurately predicts the probabilities \(p_n\) and the corresponding wave functions in both regions II and III, the experimental data deviate substantially from these expectations, raising questions about their interpretation. We summarize the following key points \mbox{and arguments:}

\begin{enumerate}
\item Theoretical Consistency
 
   The theoretical model relies on well-established quantum mechanical principles and boundary conditions for wave functions in confined regions. The wave functions \(\psi_n(z)\) and \(\tilde{\psi}_{n'}(z)\) are derived rigorously using Airy functions, with roots calculated to a high precision. Furthermore, the completeness relation ensures that the total probability is conserved. This theoretical rigor results in spatial probability distributions that are self-consistent and numerically stable, particularly for the first 100 quantum gravitational states in region III.

\item Comparison with Experimental Data

   The comparison between our theoretically derived probabilities \(p_n\) and those reported by Ichikawa {et al.} reveals agreement only for the first two quantum gravitational states. Beyond these, the experimental probabilities deviate significantly, exhibiting trends that are neither consistent with quantum mechanical predictions nor indicative of random noise. These discrepancies suggest that the experimental data interpretation may not accurately capture the quantum behavior of ultracold neutrons.

\item Unexplained Oscillations in High States
 
   The probabilities \(p_n\) for higher quantum states (\(n > 20\)) exhibit random oscillations, as predicted by our model, which are not observed in the experimental data by Ichikawa {et al.} These oscillations arise naturally from the orthogonality of wave functions and their overlap integrals, indicating a failure of the experimental setup to resolve or detect higher-order quantum gravitational states accurately.

\item Sensitivity to Initial Conditions
 
   The wave functions in region II, \(\tilde{\psi}_{n'}(z)\), are highly sensitive to the precision of the roots \(\tilde{\xi}_{n'}\). This sensitivity impacts the probabilities \(|\tilde{a}_{n'}(0)|^2\) and their subsequent propagation into region III. Our analysis shows that minute deviations in these roots could lead to notable changes in the spatial distribution in region III. However, this effect does not account for the observed experimental discrepancies, which are far larger than what could be attributed to root precision.

\item Experimental Limitations and Systematic Errors
 
   The experimental setup described by Ichikawa {et al.} might introduce systematic biases that are not adequately accounted for:
   \begin{itemize}
\item Beam Characteristics: The distribution of ultracold neutrons in the beam may not perfectly correspond to the assumed probability distribution \(P(-\tilde{\xi}_{n'}\ell_0)\), affecting the initial conditions in region II.
\item Detector Resolution: The pixelated detector used to project spatial distributions in region III may have insufficient resolution to accurately measure higher-order quantum states.
\item Uncontrolled Environmental Factors: Gravitational gradients, vibrations, or mirror imperfections in the experimental setup could distort the spatial distribution of ultracold neutrons.
\end{itemize}

\item Reevaluation of Experimental Assumptions
 
   The interpretation of Ichikawa {et al.} implicitly assumes that the spatial distributions in region III directly reflect the mixed quantum state formed in region II. However, the complex dynamics at the boundary between regions II and III may introduce additional effects, such as decoherence or scattering, that are not accounted for in the experimental analysis. Such effects could explain the observed deviations from the theoretical predictions.

\item Wigner Function Analysis
 
   As detailed in the previous section, the analysis of the Wigner function further supports the conclusion that the experimental data lack the expected quantum coherence. The Wigner function analysis reveals that the observed spatial distributions are inconsistent with the quantum mechanical predictions for mixed states of ultracold neutrons. This suggests that the experimental data may be dominated by classical or semiclassical effects, rather than pure quantum gravitational states.
   
\end{enumerate}

In light of these arguments, it is evident that the experimental results reported by Ichikawa {et al.} do not align with the spatial distributions predicted by quantum mechanics for ultracold neutrons in gravitational fields. To resolve these discrepancies, further experimental investigations are required, with particular attention to the following:
\begin{enumerate}
\item Enhanced precision in beam preparation and state initialization in region II.
\item Improved resolution and calibration of the pixelated detector in region III.
\item A detailed study of environmental influences and systematic errors.
\end{enumerate}

Additionally, theoretical models should be extended to include potential decoherence mechanisms and other perturbative effects that may arise at the interface between regions II and III. Such efforts would provide a more comprehensive understanding of the dynamics of ultracold neutrons in quantum gravitational states and bridge the gap between theory and experiment.

\section{Conclusions}
\label{sec:conclusion}

To construct a robust model for ultracold neutrons (UCNs) interacting with gravitational fields and cylindrical mirrors, it is essential to incorporate simplifying assumptions. A common assumption is that the UCN beam reaches a steady-state distribution. This simplifies the analysis by focusing on the time-independent behavior of the neutrons. Typically, the initial velocity distribution of UCNs follows a Maxwell--Boltzmann distribution at very low temperatures, as reported by \cite{Ichikawa2014}. %MDPI: Please cite all references with reference numbers and place the numbers in square brackets ("[ ]"), e.g., [1], [1-3], or [1,3]. Please refer to the following website for more information: http://www.mdpi.com/authors/references. 

The interaction of neutrons with mirror surfaces involves complex phenomena such as absorption, reflection, and scattering. To simplify the model, we assume that the mirrors are ideal, perfectly reflective surfaces. This implies that neutrons undergo specular reflection, where the angle of incidence equals the angle of reflection, with no energy loss. This assumption is valid for mirrors that are smooth and highly reflective to UCNs, which is typically achieved by using specialized coatings designed for neutron reflectivity. By adopting this idealization, we avoid complications arising from microscopic surface properties or inelastic scattering processes, making the boundary conditions tractable mathematically.

We analyzed experimental data on the spatial distribution of quantum gravitational states of UCNs obtained by Ichikawa {et al.} \cite{Ichikawa2014}. In their experimental setup, UCNs with a horizontal velocity \(v_x\), following a Gaussian distribution with a mean velocity \(v_0 = 9.4\,\mathrm{m/s}\) and variance \(\Delta v = 2.8\,\mathrm{m/s}\), pass through three distinct regions:
\begin{enumerate}
\item  A spatial region between two parallel mirrors (region II),
\item  A spatial region above a mirror (region III),
\item A detection region, where UCNs are observed.
\end{enumerate}

The cylindrical mirror projects UCNs from region III onto the detector. 

Due to the small de Broglie wavelength of the neutrons, \(\lambda_n = 42(13)\,\mathrm{nm}\), compared to both the distance between the two mirrors, \(h_{\mathrm{II}} = 100\,\upmu\mathrm{m}\), and the radius of the cylindrical mirror, \(R = 3\,\mathrm{mm}\), the scattering of UCNs by the cylindrical mirror can be treated classically. In this framework, the impact parameter \(b\) of a neutron corresponds to its vertical coordinate \(z\). For analyzing the spatial distribution of UCNs on the detector, we project the impact parameter \(b\) onto the detector variable \(Z\). For the experimental setup in \cite{Ichikawa2014}, UCNs are detected only within an interval \(\Delta Z = 2.34\,\mathrm{mm}\). The projection is defined such that UCNs with \(b = 0\) are mapped to \(Z = 0\).

To calculate the probability density distributions of quantum gravitational states of UCNs and the corresponding Wigner functions, we define the wave function of the mixed state in region II as a superposition of 15 quantum gravitational states. The expansion coefficients \(a_n\) (\(n = 1, 2, \ldots, 15\)) are determined by the \(z\)-distribution of UCNs. The wave functions of these quantum gravitational states between the two mirrors have been previously computed in \cite{Ivanov2013}. 

When expanding the wave function of UCNs from region II into the quantum gravitational states in region III, we include up to 100 states, characterized by expansion coefficients \(a_n(t_0)\) (\(n = 1, 2, \ldots, 100\)). This expansion satisfies the unitarity condition, \(\sum_{n=1}^{100}\overline{|a_n(t_0)|^2} = 0.9991\), where averaging is performed over random phases of the coefficients \(\tilde{a}_{n'}(0)\) in region II.

In Figure\,\ref{fig:DN7qa}, we show that the probabilities \(p_n\) for UCNs to occupy the \(n\)-th quantum gravitational state differ significantly from those calculated by Ichikawa {et al.} \cite{Ichikawa2014}. Specifically, the smooth dependence of \(p_n\) on \(n\) for \(n \geq 20\), reported by Ichikawa {et al.}, is inconsistent with our findings. The wave function of the mixed state is localized within \(0 \leq z \leq 100\,\upmu\mathrm{m}\), leading to an irregular and oscillatory dependence of \(p_n\) on \(n\) for \(n \geq 20\).

Our analysis also reveals that the calculated probability distributions (Figure~\ref{fig:DN8qa}) and the Wigner function (Figure~\ref{fig:DN9q}) do not align with the experimental data from Ichikawa {et~al.}~\cite{Ichikawa2014}. This discrepancy suggests that their data cannot be fully explained by the spatial distribution of quantum gravitational states of UCNs under our theoretical framework.

%R.H\"ollwieser is receiving funding from the program " Netzwerke 2021", an initiative of the Ministry of Culture and Science of the State of Northrhine Westphalia, in the NRW-FAIR network, funding code NW21-024-A. B 128/5-2. 

\vspace{6pt}
\authorcontributions{D.A.: Conceptualization, Methodology, Data Curation, Formal analysis, Investigation, Writing—Original draft preparation. R.H.: Conceptualization, Formal analysis, Data Curation, Investigation, Methodology, Software, Validation, Writing—Original draft preparation. M.W.: Conceptualization, Formal analysis, Data Curation, Investigation, Methodology, Software, Validation, Writing—Original draft preparation. All authors contributed equally and agreed to the published version of the manuscript. The sole responsibility for the content of this publication lies with the authors.}

\funding{The work of M. Wellenzohn was supported by MA 23 (p.n. 30-22).}

\dataavailability{The data and illustrations presented in this study can be obtained directly from the equations. All data are available upon request from the corresponding author.}
\acknowledgments{We want to acknowledge our dear colleague  Andrey Nikolaevich Ivanov, who was the main investigator of this work until he sadly passed away on 18 December 2021. We see it as our professional and personal duty to honor his legacy by continuing to publish our collaborative work. Andrey was born on 3 June 1945 in what was then Leningrad. Since 1993 he was a university professor at the Faculty of Physics, named “Peter The Great St. Petersburg Polytechnic University” after Peter the Great. Since 1995 he has been a guest professor at the Institute for Nuclear Physics at the Vienna University of Technology for several years and has been closely associated with the institute ever since. This is also where we met Andrey and have been collaborating with him closely over more than 20 years, resulting in 40 scientific publications, see also~\cite{Ivanov:2013fca, Ivanov2013a, Ivanov2014, Ivanov2014a, Ivanov:2017mnz, Ivanov:2017wxl, Ivanov2017b, Ivanov:2018uuk, Ivanov:2018vit, Ivanov:2018vmz, Ivanov:2018qen, Ivanov:2018ngi, Ivanov:2018olo, Ivanov:2018yir, Ivanov:2019rkp, Ivanov:2019bqr, Ivanov:2020ybx, Ivanov:2021bae, Ivanov:2021lji, Ivanov:2021yhl, Ivanov:2021xkm, Altarawneh:2024lsf, Altarawneh:2024sxo}. We will miss Andrey as a personal friend for his immense wealth of ideas, scientific skills, and  creativity. See also  the {official obituary} for Andrey Nikolaevich Ivanov. The work of M. Wellenzohn was supported by MA 23 (p.n. 27-07 and p.n. 30-22). The sole responsibility for the content of this publication lies with the authors.}
\conflictsofinterest{The authors declare no conflicts of interest.}

\begin{adjustwidth}{-\extralength}{0cm}

\reftitle{References}

\PublishersNote{}
\end{adjustwidth}
\end{document}